\documentclass[fleqn,usenatbib]{mnras}
\usepackage{amsmath,amstext,mathrsfs}
\usepackage{txfonts}
\usepackage{enumitem}
\usepackage{float}
\usepackage{color} 
\usepackage{natbib}
\usepackage{subcaption}
\usepackage{gensymb}
\usepackage{natbib}
\usepackage{amssymb}
\usepackage{bm}
\usepackage{rotating}
\usepackage{graphicx,float}
\usepackage{caption}
\usepackage{tabularx}
\usepackage{longtable}
\usepackage{xcolor}
\usepackage{graphicx}
\usepackage{tikz}
\usepackage{soul}
\usepackage{footmisc}
\usepackage{comment}
\usepackage{makecell}
\usepackage{multirow}
\usepackage{ulem}
\usepackage[T1]{fontenc}

\usepackage{scalerel}
\usepackage{tikz}
\usetikzlibrary{svg.path}
\definecolor{orcidlogocol}{HTML}{A6CE39}
\tikzset{
	orcidlogo/.pic={
		\fill[orcidlogocol] svg{M256,128c0,70.7-57.3,128-128,128C57.3,256,0,198.7,0,128C0,57.3,57.3,0,128,0C198.7,0,256,57.3,256,128z};
		\fill[white] svg{M86.3,186.2H70.9V79.1h15.4v48.4V186.2z}
		svg{M108.9,79.1h41.6c39.6,0,57,28.3,57,53.6c0,27.5-21.5,53.6-56.8,53.6h-41.8V79.1z M124.3,172.4h24.5c34.9,0,42.9-26.5,42.9-39.7c0-21.5-13.7-39.7-43.7-39.7h-23.7V172.4z}
		svg{M88.7,56.8c0,5.5-4.5,10.1-10.1,10.1c-5.6,0-10.1-4.6-10.1-10.1c0-5.6,4.5-10.1,10.1-10.1C84.2,46.7,88.7,51.3,88.7,56.8z};
	}
}

\newcommand\orcidicon[1]{\href{https://orcid.org/#1}{\mbox{\scalerel*{
				\begin{tikzpicture}[yscale=-1,transform shape]
					\pic{orcidlogo};
				\end{tikzpicture}
			}{|}}}}

\DeclareRobustCommand{\VAN}[3]{#2}
\let\VANthebibliography\thebibliography
\def\thebibliography{\DeclareRobustCommand{\VAN}[3]{##3}\VANthebibliography}

\title[GZ Effect: Fast Radio Bursts]{Gertsenshtein-Zel$'$dovich effect: A plausible explanation for fast radio bursts?}
\author[Kushwaha et al.]{
Ashu Kushwaha
\orcidicon{0000-0001-9910-5010}$^{1}$
\thanks{E-mail:ashu712@iitb.ac.in (AK)},
Sunil Malik
\orcidicon{0000-0003-4147-626X}$^{1,2,3}$
\thanks{sunil.malik@uni-potsdam.de (SM) (Corresponding Author) },
S. Shankaranarayanan
\orcidicon{0000-0003-2560-8066}$^{1}$
\thanks{shanki@phy.iitb.ac.in}
\\
$^{1}$Department of Physics, Indian Institute of Technology Bombay, Mumbai 400076, India\\
$^{2}$Institute fur Physik und Astronomie Universitat Potsdam, Golm Haus 28, D-14476 Potsdam, Germany\\
$^{3}$Deutsches Elektronen-Synchrotron DESY, Platanenallee 6, 15738 Zeuthen, Germany
}
	
\date{Accepted XXX. Received YYY; in original form ZZZ}

\begin{document}

\label{firstpage}
\maketitle
 
\begin{abstract}
We present a novel model that may provide an interpretation for a class of non-repeating FRBs --- short ($<1~\rm{s}$), bright ($0.1 - 1000~\rm{Jy}$) bursts of MHz-GHz frequency radio waves. The model has three ingredients --- compact object, a progenitor with effective magnetic field strength around $10^{10}~{\rm Gauss}$, and high frequency (MHz-GHz) gravitational waves (GWs). At resonance, the energy conversion from GWs to electromagnetic waves occurs when GWs pass through the magnetosphere of such compact objects due to the Gertsenshtein-Zel'dovich effect.
This conversion produces bursts of electromagnetic waves in the MHz-GHz range, leading to FRBs. Our model has three key features:  (i) predict peak-flux, (ii) can naturally explain the pulse width, and  (iii) coherent nature of FRB. We thus conclude that the neutron star/magnetar could be the progenitor of FRBs. Further, our model offers a novel perspective on the indirection detection of GWs at high-frequency beyond detection capabilities. Thus, transient events like FRBs are a rich source for the current era of multi-messenger astronomy.
\end{abstract}
\begin{keywords}
Transient phenomena, Fast Radio Bursts, Neutron stars, Pulsars, magnetic fields, Gravitational Waves
\end{keywords}

\section{Introduction}
Technological advancement has fuelled research in high-energy astrophysical phenomena at larger redshift ranges, and we are in a position to address some unresolved questions starting from pulsar emission mechanism~\cite{2021MNRAS_pulsar} to short bursts such as Gamma-ray bursts (GRBs)~\cite{2016-Levan.etal-SSR}, Fast radio bursts (FRBs)~\cite{2008-Lorimer-LivRevRel,2019-Cordes.Chatterjee-AnnRevAA,2019-Platts.etal-PhyRept}. To date, more than 600 FRBs have been reported in various catalogues~\cite{Petroff:2016tcr,2019-Platts.etal-PhyRept, Pastor-Marazuela:2020tii,2021-Rafiei.etal-APJ}. $99\%$ of these FRBs have the following three characteristic features: observed peak flux ($S_{\nu}$) 
varies in the range $0.1~{\rm Jy} < S_{\nu} < 700~{\rm Jy}$, coherent radiation and the pulse width is less than a second \cite{2021-Rafiei.etal-APJ,Petroff:2016tcr}. These observations have posed the following questions: What causes these extreme high-energy transient radio-bursts from distant galaxies, lasting only a few milliseconds each~\cite{2008-Lorimer-LivRevRel,2019-Cordes.Chatterjee-AnnRevAA,2019-Platts.etal-PhyRept}? Why do some FRBs repeat at unpredictable intervals, but most do not~\cite{2019-Platts.etal-PhyRept}?
Does strong gravity provide an active role? 

 {Naturally, many models have been proposed to explain the origin of FRBs. All these models try to provide a physical mechanism that results in large amount of coherent radiation in a short time~\cite{2008-Lorimer-LivRevRel,2019-Cordes.Chatterjee-AnnRevAA,Petroff:2016tcr,2019-Platts.etal-PhyRept, Pastor-Marazuela:2020tii,2021-Rafiei.etal-APJ}. Since the time scale of these events is less than a second, and the emission is coherent, the astrophysical processes that explain these events \emph{cannot} be thermal~\cite{2008-Lorimer-LivRevRel}.}

 {Broadly, these models can be classified into two categories~\cite{2022-Zhang-arXiv}: FRBs created by interaction of an object with a pulsar/magnetar and FRBs created from the magnetar/pulsar itself~\cite{2008-Lorimer-LivRevRel,2019-Cordes.Chatterjee-AnnRevAA,Petroff:2016tcr,2019-Platts.etal-PhyRept, Pastor-Marazuela:2020tii,2021-Rafiei.etal-APJ}. The first category can further be classified into two broad classes. In the first class, the energy powering FRBs comes from
the neutron star magnetosphere/wind themselves, and an orbiting object converts this energy into
radiation. In the second class, the object falls onto the neutron star, and its gravitational energy partly gets converted to FRBs.
In the second class, many models involving non-thermal processes such as Synchrotron radiation~\cite{Book-Lorimer.Kramer-PulsarAstronomy}, black hole super-radiance~\cite{2018-Conlon.Herdeiro-PLB}, evaporating primordial black hole \cite{1977-Rees-Nature,2020-Carr.Kuhnel}, spark from cosmic strings~\cite{2008-Vachaspati-PRL}, Quark Novae~\cite{2015-Shand.etal-RAA},  {synchrotron maser shock model~\cite{2020ApJ...900L..26W},  radiation from reconnecting current
sheets in the far magnetosphere~\cite{2020ApJ...897....1L},  curvature emission from charge bunches~\cite{2022ApJ...927..105W} } have been proposed. Several classes of FRB models predict prompt multiwavelength counterparts and
specify the ratio between the energy emitted by the counterpart
and by the FRB~\cite{2017-Zhang-ApJL,2019-Metzger.etal-MNRAS}.}

However, despite the use of exotic new physics, no single model has provided a universal explanation for the enormous energy released in these events. It is important to note that all these mechanisms require  electromagnetic interaction to generate FRBs.
Due to the nature of electromagnetic interaction, small-scale emission mechanisms usually predominate over large-scale coherent electromagnetic processes (like astrophysical masers and pulsar radio emission). In this work, we provide an alternative framework that overcomes this and can explain the observed coherence in FRBs.

As shown below, one key missing ingredient is the \emph{dynamics of strong-gravity}. The Spatio-temporal changes in the strong-gravity regime --- oscillons,  phase transitions, plasma instability, primordial black holes, reheating --- generate gravitational waves (GWs) in a broad range of frequencies ($10^{-15} - 10^{15}$ Hz) \cite{1974-Hawking.Carr-MNRAS,2009PhRvL.103k1303A,2015-Kuroda.etal-IJMPD,2019-Ejlli.etal-EPJC,2020-Aggarwal.etal-arXiv,2020-Chen.Rajendran.etal-arXiv,2021-Pustovoit.etal-JOP}. Like EM waves,  {GWs are generated by the time-varying quadrupole moment}~\cite{2009-Sathyaprakash.Schutz-LivRevRel,2000-Schutz-arXiv}. Since all masses have the same gravitational sign and tend to clump together, they produce large coherent bulk motions that generate \emph{energetic, coherent GWs}~\cite{2007-Hendry.Woan-AG}. Thus, if a mechanism that converts incoming coherent GWs to EM waves exists, we can explain the extremely energetic, coherent nature of FRBs~\cite{2018-Popov.etal-Usp,2022-Lieu.etal-CQG}. In this work, we construct a model that uses this feature. 

 {Since FRBs are highly energetic, an attentive reader might wonder do incoming GWs carry such large energies.} GWs carry an enormous amount of energy. For example, typical GWs from a compact binary collapse with amplitude $h \sim 10^{-22}$ carry the energy of the order of $10^{20}~\rm{Jy}$ \cite{2009-Sathyaprakash.Schutz-LivRevRel,2000-Schutz-arXiv}.  {Also, stellar mass binary black hole collision can lead to
peak gravitational-wave luminosity of
$10^{56}~\rm{ergs/s}$~\cite{Book-Schutz,2016-LIGOScientific-PRL}.} If the GWs indeed carry a lot of energy, can this energy transform into other observable forms of energy?  
Currently, there is evidence of GWs in the frequency range $10^{-9} - 10^{4}~{\rm Hz}$ from LIGO-VIRGO-KAGRA and PTA observations~\cite{2016-Abtott.etal_LIGOScientific-PRL,2023-Agazie.etal-AAS}. While most of the current effort has focused on these frequency ranges, there is a surge in activity for the possibility of detecting GWs in the MHz-GHz frequency range~\cite{2020-Aggarwal.etal-arXiv}. New physics beyond the standard model of particle physics, like an exotic compact object, can produce observable GW signals in this frequency ranges~\cite{2020-Aggarwal.etal-arXiv,2020-Chen.Rajendran.etal-arXiv,2021-Pustovoit.etal-JOP}.

\begin{figure*}
\centering
\includegraphics[height=1.6in]{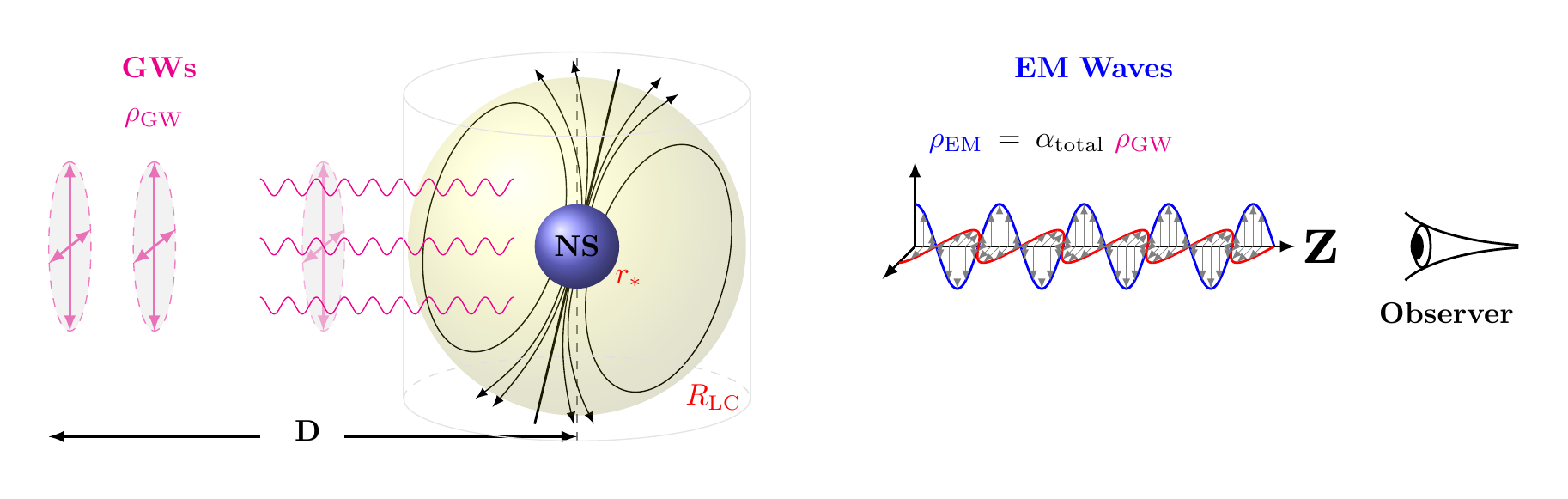}%
\caption{Schematic depiction of GZ effect. The externally generated GWs at distance $D$ from the NS is converted to EM Waves in the magnetosphere.The blue region corresponds to NS with radius $r_*$ and the yellow region around NS corresponds to magnetosphere of radius $R_{\rm LC}$. The black curves correspond to the magnetic field lines.}
\label{fig:setupFigure}
\end{figure*}

A physics maxim is that energy can be transformed between different forms. Although the total energy is conserved, the efficiency of the transformation depends on the energy scale, background dynamics, and external conditions (parameters). Energy transformation is one way to probe strong gravity regions like the early universe, black-holes, and NS. In this work, we propose a novel approach that uses the energy conversion from incoming, coherent GWs to electromagnetic (EM) waves that can \emph{explain milli-second bursts, like non-repeating FRBs}.

GWs get converted to EM waves in the presence of strong transverse magnetic fields --- Gertsenshtein-Zel'dovich (GZ) effect~\cite{1962-Gertsenshtein-JETP,1974-Zeldovich-SJETP,2018-Zheng.Wei.Li-PRD,2021-Domcke.Garcia-Cely-PRL}. 
Gertsenshtein postulated the existence of wave resonance between EM waves and GWs on the basis of their identical propagation velocities and the equations describing them are linear. By employing linearized Einstein's field equations, he demonstrated that EM waves are generated through wave resonance when gravitational waves traverse a strong magnetic field. In the same way, light passing through a strong magnetic field generates gravitational waves~\cite{1974-Zeldovich-SJETP,2015-Kolosnitsyn-Phys.Scripta}. Appendix  \ref{appsec:E&B-solution} contains the detailed calculations reproducing the key results.

To understand the GZ effect, consider coherent GWs with a frequency $\omega_g$ passing through a region with a high transverse static magnetic field ${\bf B}$. The propagation of GWs leads to compression and stretching of the magnetic field proportional to $h {\bf B}$ ($h$ is the amplitude of GWs), which acts as a source leading to the generation of EM waves ~\cite{1962-Gertsenshtein-JETP,1974-Zeldovich-SJETP}. The induced (resultant) EM waves generated will have maximum amplitude at resonance, i.e., the frequency of EM waves is identical to $\omega_g$. (For details, see appendix~\ref{appsec:E&B-solution}) 

 {In quantum mechanical language, the GZ effect is analogous to the mixing of neutrino flavors --- the external field \emph{catalyzes} a resonant mixture of photon and graviton states~\cite{2023-Palessandro.Rothman-PDU}. The external magnetic field provides the extra angular momentum necessary for the spin-1 (photon) field to mix with the spin-2 (graviton) field. Thus, the GZ mechanism involves the transfer of energy from the incoming, coherent GWs to emitted, coherent EM radiation in the presence of the background magnetic field. 
The maximum efficiency of this conversion can be achieved if the background field is strong at the resonance frequency (when the frequency of the emitted EM radiation is the same as the incoming GWs).
Hence, the background magnetic field acts as a catalyst in this mechanism.}

We show that the {GZ effect} may provide an interpretation for a class of \emph{non-repeating FRBs}. 
To explain the energy bursts in FRBs, we propose a model with the following two realistic assumptions: 
(i) the astrophysical object is compact and has a strong gravity environment, and 
(ii) the object possesses a small time-dependent magnetic field on top of the large, effective static, transverse magnetic field. 
These two assumptions principally lead us to stellar remnants, such as NS and magnetars with magnetic field strength ranging from $10^{8} - 10^{15}~\rm{G}$~\cite{2020-Andersen.etal.CHIME-FRB-Nature,2015-Belvedere.etal-ApJ,2019-Cordes.Chatterjee-AnnRevAA,2008-Lorimer-LivRevRel}. 
The small time-dependent magnetic fields arise due to the rotation of the NS about its axis with frequency $\omega_B$~\cite{2015-Belvedere.etal-ApJ,2008-Melikidze.gil-ProcIAU,2019-Platts.etal-PhyRept,2019-Pons.Vigan-arXiv,2012-Pons.etal-AandA}. We consider $\omega_B$ in the range $[1, 10^3]~{\rm Hz}$~\cite{Book-Lorimer.Kramer-PulsarAstronomy}. 
As a result, the effective magnetic field at a given point in the NS magnetosphere is ${\bf B}(t) = {\bf B}^{(0)} + \delta {\bf B} \sin(\omega_B t)$. It has been noted that $|\delta {\bf B}/{\bf B}^{(0)}|$ can be as large as $0.1$~\cite{2012-Pons.etal-AandA}. Here, we take $|\delta {\bf B}/{\bf B}^{(0)}| \sim  10^{-2}$.

\section{Model}

\ref{fig:setupFigure} gives the schematic depiction of the physical model to explain the energy burst in FRBs. Consider GWs generated due to exotic compact objects (such as Boson stars, Oscillons, gravastars)~\cite{2020-Aggarwal.etal-arXiv,2020-Chen.Rajendran.etal-arXiv,2021-Pustovoit.etal-JOP} passing through the magnetosphere of NS 
at a distance $D$. In the figure, the magnetosphere is depicted as a cylinder. GZ effect converts GWs to EM waves as they pass through the magnetosphere~\cite{1962-Gertsenshtein-JETP,1974-Zeldovich-SJETP}. This conversion occurs at all points in the magnetosphere. 
Therefore, a faraway observer will see the integrated effect happening in the entire magnetosphere in this short duration.
For example, the light cylinder radius ($R_{\rm LC}$) for a typical NS is $\sim 10^{7}-10^{9}~{\rm cm}$, implying that the GWs take less than one second to cover the entire magnetosphere. This is one of the primary ingredients supporting our analysis for the FRB observations. 

To compute the GZ effect at a point in the magnetosphere, we consider source-free Maxwell's equations on the background space-time with GW fluctuations (see details in Appendix~\eqref{appsec:E&B-solution}). The linearized Einstein's equations, up to first order in the space-time 
perturbations are highly accurate. In this approximation, the effects of GWs on the stress-tensor (Riemann tensor) are negligible. Hence, we consider background space-time to be Minkowski (in cartesian coordinates)~\cite{Book-Gravitation_MTW}). 
The two polarizations of GW (with frequency $\omega_g$ and wave-vector $k_g$) propagating along the z-direction are:
\begin{align}\label{eq:h-Expression}
h_+  = 
A_+ \, e^{i \left( k_g z - \omega_g t \right) }, 
h_{\times}  = 
i A_{\times} \, e^{i \left( k_g z - \omega_g t \right) } ,
\end{align}
where $A_+$ and $A_{\times}$ are the constant amplitudes of the GWs. We assume that both the modes of GWs are generated with an equal amount of energy, i.e., $|A_{+}| = |A_{\times}| $  --- the isospectrality condition in general relativity~\cite{Chandrasekhar_BlackHoles-Book}.
Taking the distance between coherent GW source, such as an exotic compact object, and NS to be $D = 1 \,  \rm{kpc}$, gives $h \simeq A_{+} = 10^{-20}$ at $1 \, \rm{GHz}$ (see, \cite{2020-Aggarwal.etal-arXiv} and Appendix \ref{appsec:ECO}). In this work, we have assumed $h$ to be 
three orders smaller ($\sim 10^{-23}$) near the NS.

As mentioned above, the key requirement of the GZ-effect is the presence of the transverse magnetic field to the direction of propagation of coherent GWs. Besides, the time taken by the GWs to pass through the entire magnetosphere is much smaller than the rotation period of the millisecond pulsar/magnetar ($\omega_B^{-1}$). Given the direction of propagation of GWs along the z-axis, the effective time-dependent transverse magnetic field is taken to be
$\textbf{B}(t) = \left( 0, B^{(0)}_y + \delta B_y \sin (\omega_B t), 0  \right)$
~\cite{2019-Pons.Vigan-arXiv,2012-Pons.etal-AandA}. 

Although the effective magnetic field depends on the distance from the surface of the object~\cite{2008-Lorimer-LivRevRel,2015-Belvedere.etal-ApJ,2019-Cordes.Chatterjee-AnnRevAA,2020-Andersen.etal.CHIME-FRB-Nature}, we assume that ${\bf B}(t)$ is independent of the distance from the surface up to 
$R_{\rm LC }$~\cite{Book-Lorimer.Kramer-PulsarAstronomy}.
 {In appendix~(\ref{appsec:Integration-2}), we explicitly show that the above assumption that the background magnetic field can be treated as a constant in the entire magnetosphere gives \emph{identical results} to that of the background field decreasing radially, i. e.,
\begin{align}
\left(  B_r , B_{\theta} , B_{\phi}  \right) = B_{*} \, \left( \frac{r_* }{r} \right)^3 \, \left(  2 \cos\theta, \sin\theta , 0  \right)
\end{align}
where $B_*$ is the magnetic field on the NS surface. More specifically, assuming that the NS magnetic field is dipolar, we show in appendix~(\ref{appsec:Integration-2}) that we can approximate the average magnetic field at any point in the magnetosphere to be constant. In other words, the total conversion factor we obtain using the above assumption mimics the realistic NS regions.}

Given the above setup, we now evaluate the GZ-effect in the magnetosphere of the NS and compare it with observational quantities in two steps: 
\begin{enumerate}
\item  First step involves evaluating the conversion from coherent GWs to EM waves at a typical point inside the magnetosphere. This is referred to as \emph{conversion factor} ($\alpha$). We then obtain the total conversion factor ($\alpha_{\rm tot}$) inside the entire magnetosphere \emph{at resonance} (the frequency of EM waves is identical to $\omega_g$).  {This conversion factor includes only those contributions that are along the direction of line-of-sight and coinciding with the incoming GWs. Also, only the transverse magnetic field component to the direction of propagation at each point of the magnetosphere will contribute to the emitted EM waves. This can potentially explain the coherent nature of FRBs~\cite{2013-Katz-PRD,2017-Kumar.etal-MNRAS}.} 
\item For a given conversion factor, we obtain the Poynting vector of the resultant EM waves along the direction of propagation ($S_z$). Then, we compare the theoretically derived Poynting vector with the observation of peak flux with the reported FRBs.
\end{enumerate}
The Poynting vector is a well-defined quantity for photons that travel from the source to the observer without any hindrance~\cite{Book-Carroll.Ostlie-CUP,Book-Condon.Ransom-PUP,Book-Zhang-GRB-CUP}. More specifically, assuming there is no absorption/emission of the photons during the entire journey, the Poynting vector (of the EM waves) $S_z$ is conserved (independent of the distance between the magnetosphere and observer) and is valid for emissions from compact sources. Hence, the Poynting vector estimated at a small angle (along the direction of the incoming gravitational waves) remains the same at the source and the detector. Furthermore, as shown in \ref{fig:setupFigure}, the incoming, coherent GW is along the $z-$axis in the entire magnetosphere. Therefore, the cumulative effect of the emitted EM waves is in the same direction.

To evaluate the conversion factor, solving the linearized {covariant Maxwell's} equations leads to the following electric and magnetic fields induced due to GWs, i.e., $\tilde{E}_x$ and $\tilde{B}_y$ as: 
\begin{align}\label{eq:E_x-Final}
\tilde{E}_x &\simeq  - \frac{A_{+} }{4} \, B^{(0)}_y \left( 1   -  \xi \,  \omega_B t  \,\, \right) \,\, e^{i \left(k_g z - \omega_g t\right) } 
\\
\label{eq:B_y-Final}
\tilde{B}_y &\simeq  - \frac{A_{+} }{4} \, B^{(0)}_y \left(1  +  2 \xi \,  \omega_g t  \,\,\right) e^{i \left(k_g z - \omega_g t\right) } \, ,
\end{align}
where $\xi \equiv \delta B_y/B_y^{(0)}$. Note that 
the amplitude of $\tilde{B}_y$ has a dependence on $\omega_g$, while the amplitude of $\tilde{E}_x$ has $\omega_B$ dependence on $\omega_g$. This is because the induced electric field arises due to the time-varying magnetic field (see appendix~\ref{appsec:E&B-solution}).

The conversion factor ($\alpha$) --- ratio of the energy density of EM wave and GWs --- gives the efficiency of the process at resonance. $\alpha$ for this process is 
\begin{align}\label{eq:alpha-at-point}
\!\!\! \alpha \equiv \frac{ \rho_{\rm EM } }{\rho_{\rm GW }} 
\simeq \frac{ G |B^{(0)}_y|^2 }{4 c^2 } 
\left[2   \left( \frac{\xi z}{c}\right)^2 + 2\frac{\xi}{\omega_g} \frac{z}{c} 
+ \frac{1 }{\omega_g^2} \right] 
\end{align}
where $z$ refers to the radial distance in the magnetosphere. For details, see appendix~\ref{appsec:Integration}.

\begin{figure}
\centering
\label{fig:alphaTot_magnt}%
\includegraphics[height=2.2in]{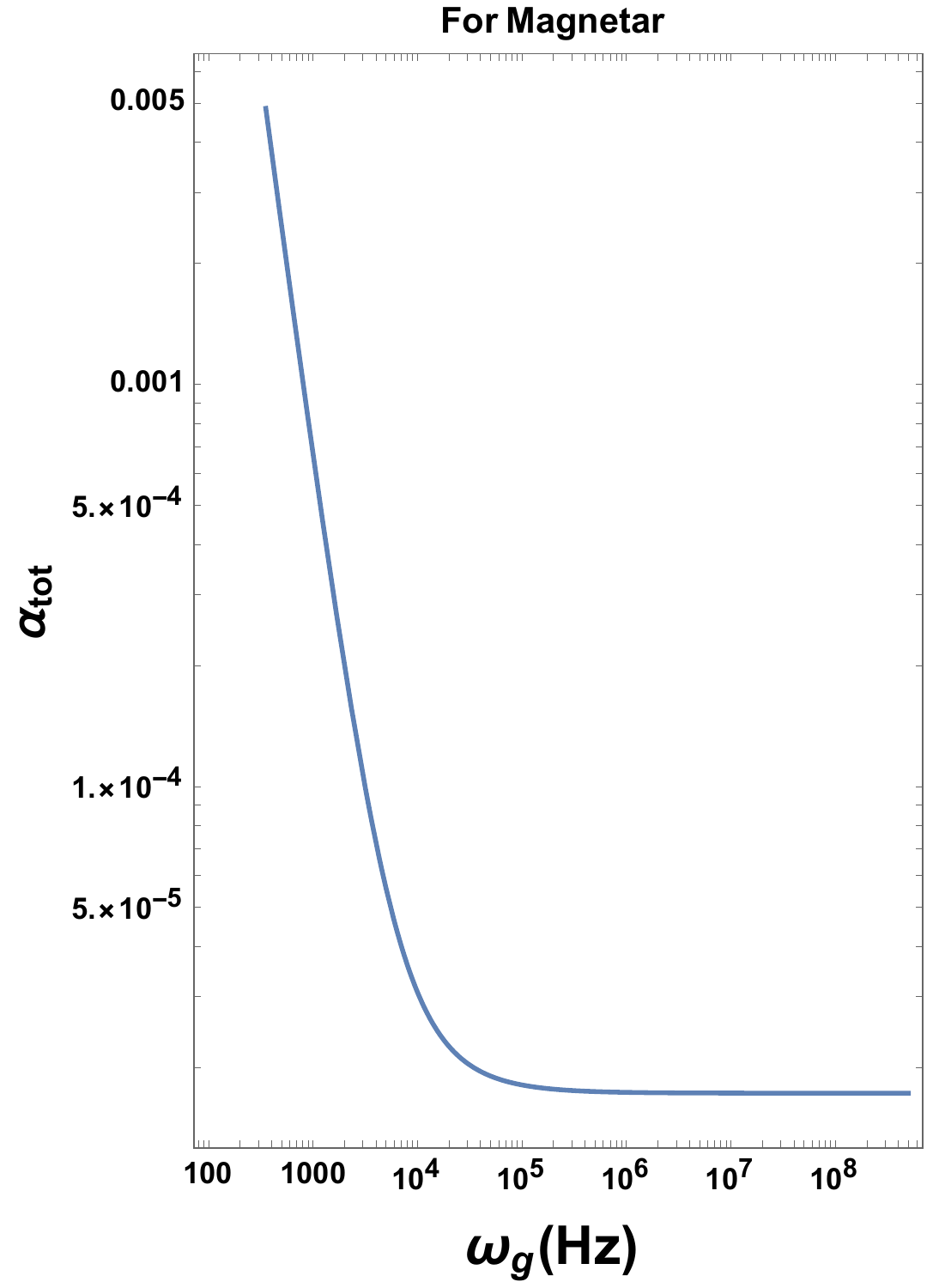} 
\label{fig:alphaTot_ns}%
\includegraphics[height=2.2in]{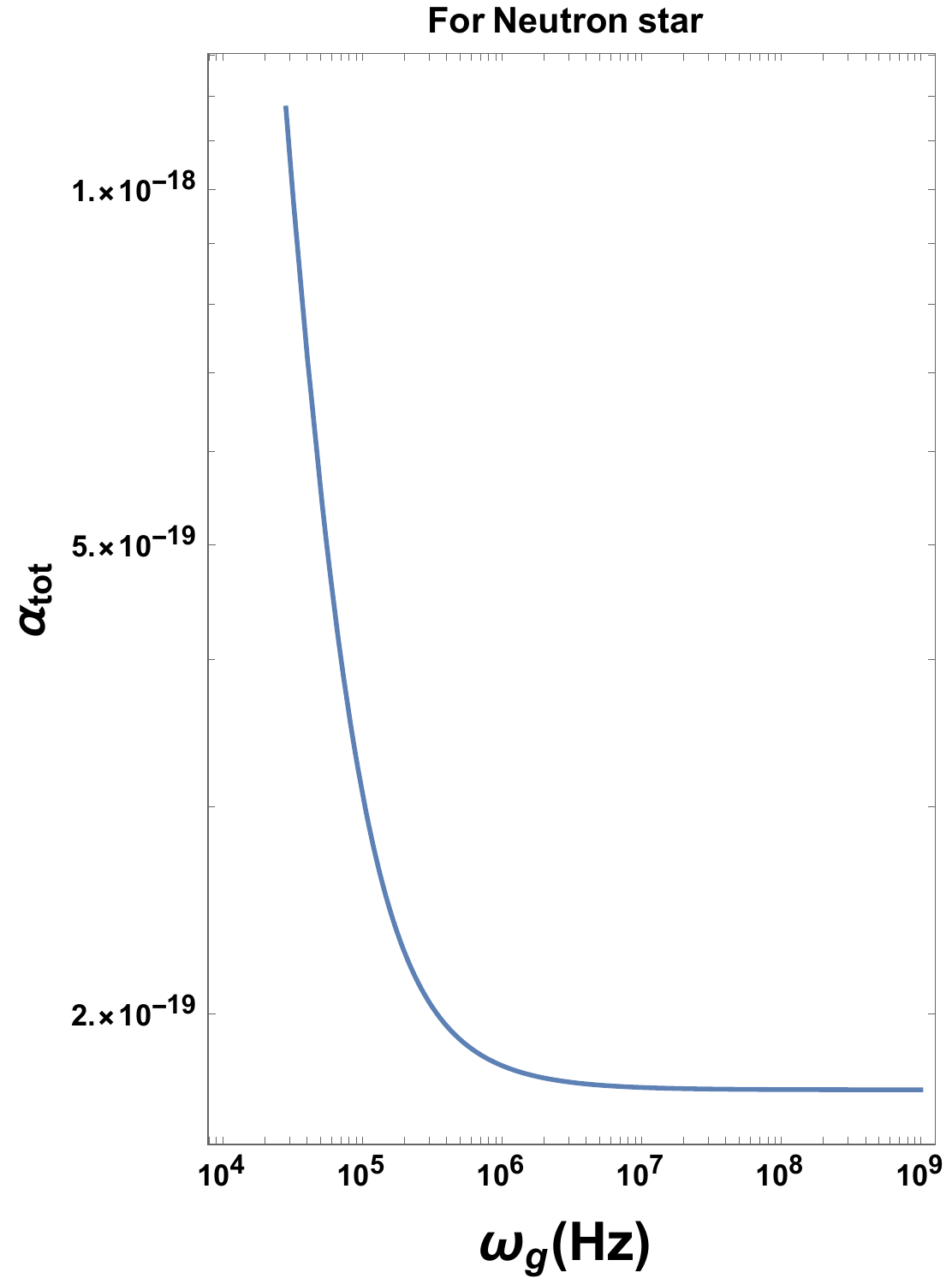} 
\caption{Log-Log plot of $\alpha_{\rm tot}$ versus $\omega_g$ for typical magnetar (left plot) and typical NS (right plot). For magnetar, we set $B^{(0)}_y = 10^{15} \, {\rm G}, R_{\rm LC} = 10^9~ \, {\rm cm},  \omega_B = 1 \, {\rm Hz}$. For NS, we set $B^{(0)}_y = 10^{10}  {\rm G} , R_{\rm LC} = 10^7 {\rm cm}  , \omega_B = 1  {\rm kHz}$.}
\label{fig:Plot-convfact}
\end{figure}

The above expression is the conversion factor at a single point on the magnetosphere. Assuming that there is no cross-correlation of GZ-effect at two distinct points in the magnetosphere, we obtain the total conversion by integrating over the entire magnetosphere (from the surface of the compact object to the light cylinder $R_{\rm LC}$). (The cross-correlation corresponds to the induced EM waves at two distinct points affecting each other. This physically corresponds to higher-order effects 
($A_{+} \partial_{z} \tilde{B}_y$) which are neglected.). Thus, the total conversion factor is 
{\small
\begin{align}\label{eq:alpha_tot}
\alpha_{\rm{tot} } \simeq
\frac{\pi G |B^{(0)}_y|^2 }{ c^2} \left[  \frac{2}{3} \left[\frac{ \xi R_{\rm LC}}{c} \right]^2 \!\! + \frac{ \xi R_{\rm LC}}{ \omega_g c} + \frac{1}{\omega_g^2} \right].
\end{align}
}
See appendix~\ref{appsec:Integration} for details. It is important to note that $\alpha_{\rm tot}$ is \emph{independent of the amplitude} of GWs. 
To understand the variation of $\alpha_{\rm tot}$ with $\omega_g$, in \ref{fig:Plot-convfact}, we have plotted the conversion factor for 
magnetar 
and NS/milli-second pulsar.

%
%

\begin{figure}
\centering
\includegraphics[height=1.6in]{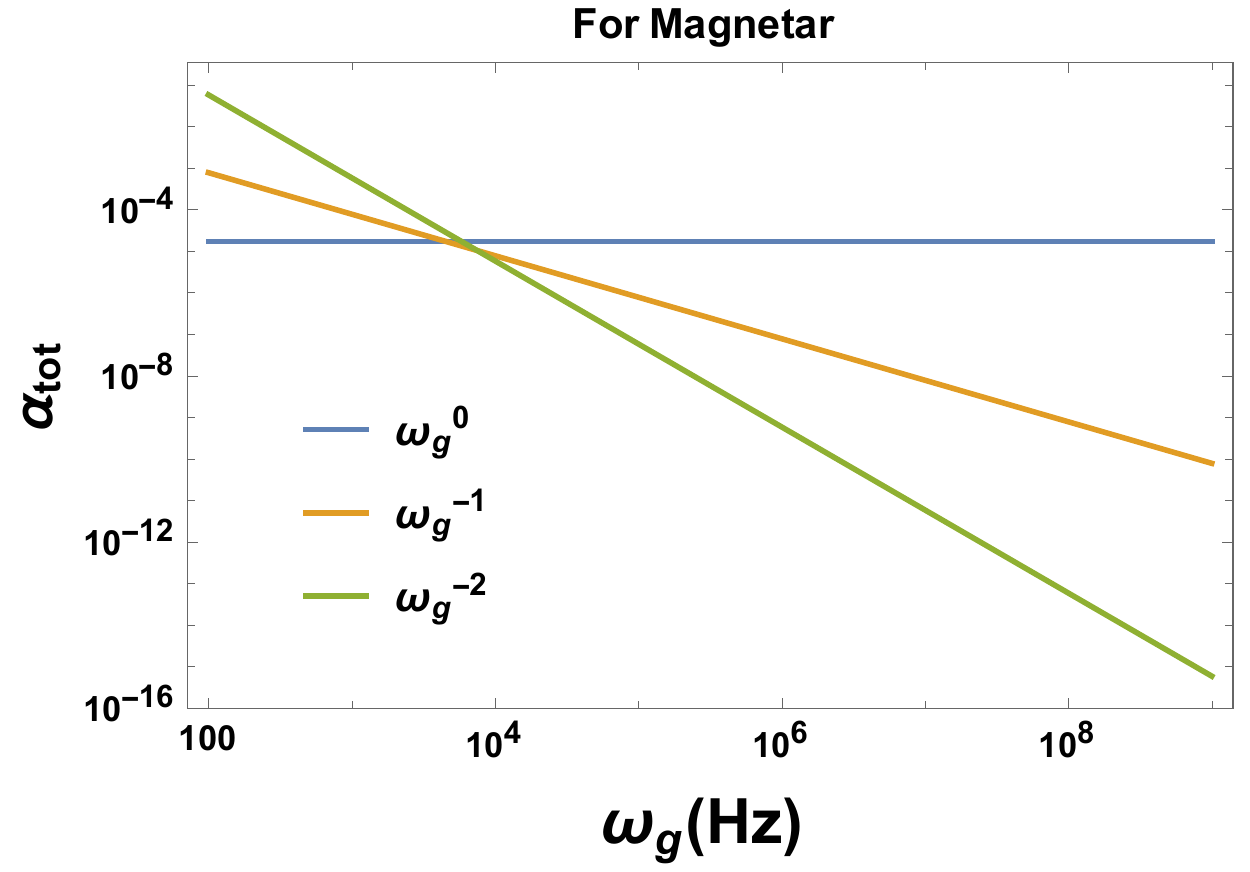} %
\centering
\includegraphics[height=1.6in]{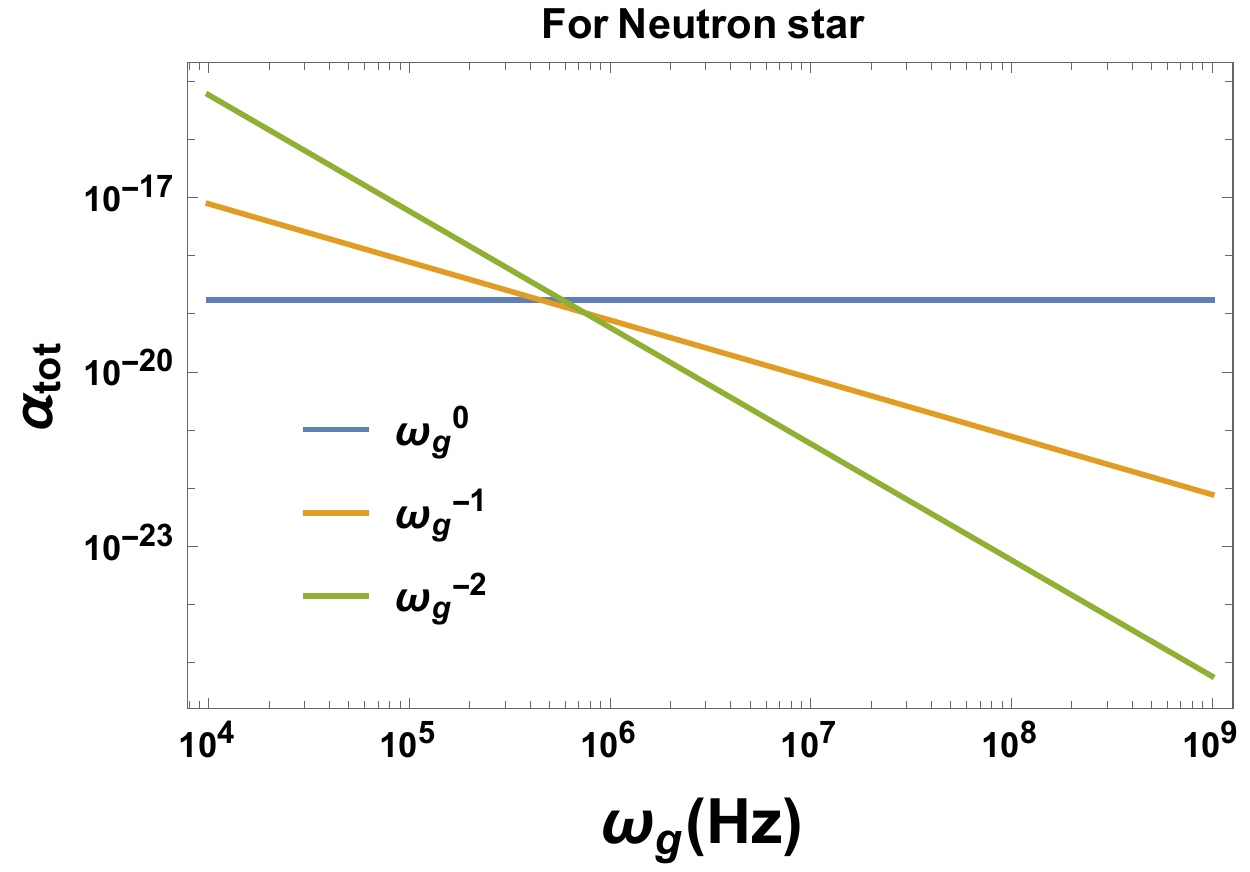} %
\caption{Log-Log plot of the three terms in the RHS of Eq.~(\ref{eq:alpha_tot}) versus $\omega_g$ for typical magnetar (top plot) and typical NS (bottom plot). For magnetar, we have set $B^{(0)}_y = 10^{15} {\rm G}, R_{\rm LC} = 10^9~ {\rm cm},  \omega_B = 1 {\rm Hz}$.
For NS/milli-second pulsar, we have set $B^{(0)}_y = 10^{10} {\rm G}, R_{\rm LC} = 10^7 {\rm cm}, \omega_B = 1 {\rm kHz}$.}
\label{fig:Plot-TermsCompare}
\end{figure}

From the plots, we infer that the total conversion factor is almost insensitive at higher frequencies ($> 1~{\rm MHz}$) for both types of compact objects. 
This is because the second and third terms in RHS of Eq.~\eqref{eq:alpha_tot} are inversely proportional to $\omega_g$ and, hence, 
the contribution to $\alpha_{\rm tot}$ is only from the first term. To elaborate, \ref{fig:Plot-TermsCompare} plots each of these terms in Eq.~\eqref{eq:alpha_tot} which 
shows a clear distinction between milli-second Pulsar and magnetar. In both cases, the cross-over occurs below 
$1~{\rm MHz}$. Since we are interested in radio frequency in the MHz-GHz range, the total conversion factor \eqref{eq:alpha_tot} is independent of the incoming, coherent GW frequency.

This leads to the important question: What is the efficiency of the GZ-effect near magnetar and NS? Table \eqref{table1} lists the total conversion factor (4th column) for magnetar and NS for different frequencies. We want to emphasize the following points: First, as mentioned earlier, we have assumed the amplitude of the GWs, i.e., $A_+ = A_\times =  1.4\times 10^{-23}$~\cite{2015-Kuroda.etal-IJMPD,2020-Aggarwal.etal-arXiv}. However, as we show below, even with this conservative value, the model explains the peak flux of FRBs. Second, $\alpha_{\rm tot}$ is very high near the magnetar for low-frequency GWs. Specifically, the conversion is $0.06 \% $ and $0.25 \%$ at 1000 Hz and 500 Hz, respectively (for the magnetar $B_y^{(0)} = 10^{15} \, {\rm G}$). For high-frequency GWs, as we show below, even a total conversion factor $(\alpha_{\rm tot})$ of $10^{-19}$ can lead to appreciable energy in the radio frequency (MHz-GHz range). 

\begin{figure}
\centering
\includegraphics[height=1.6in]{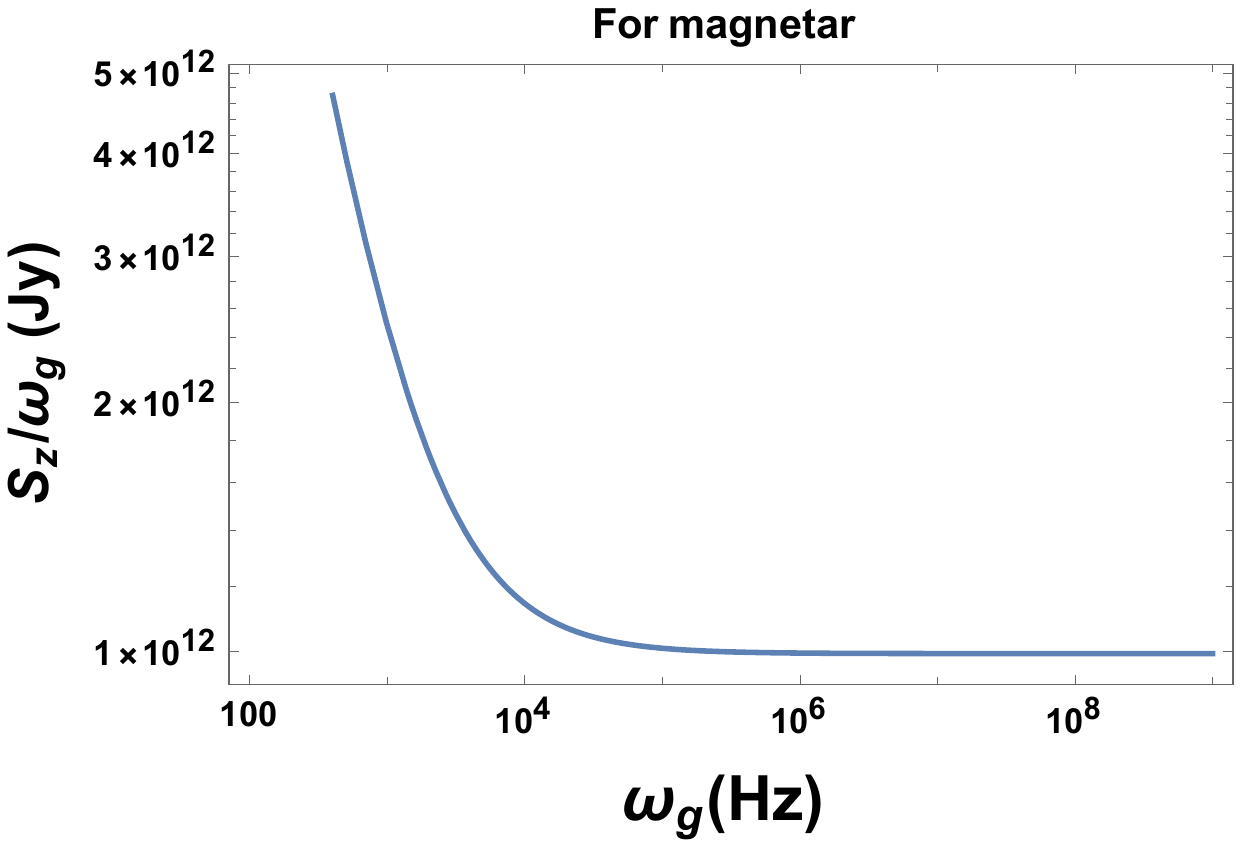} %
\centering
\includegraphics[height=1.6in]{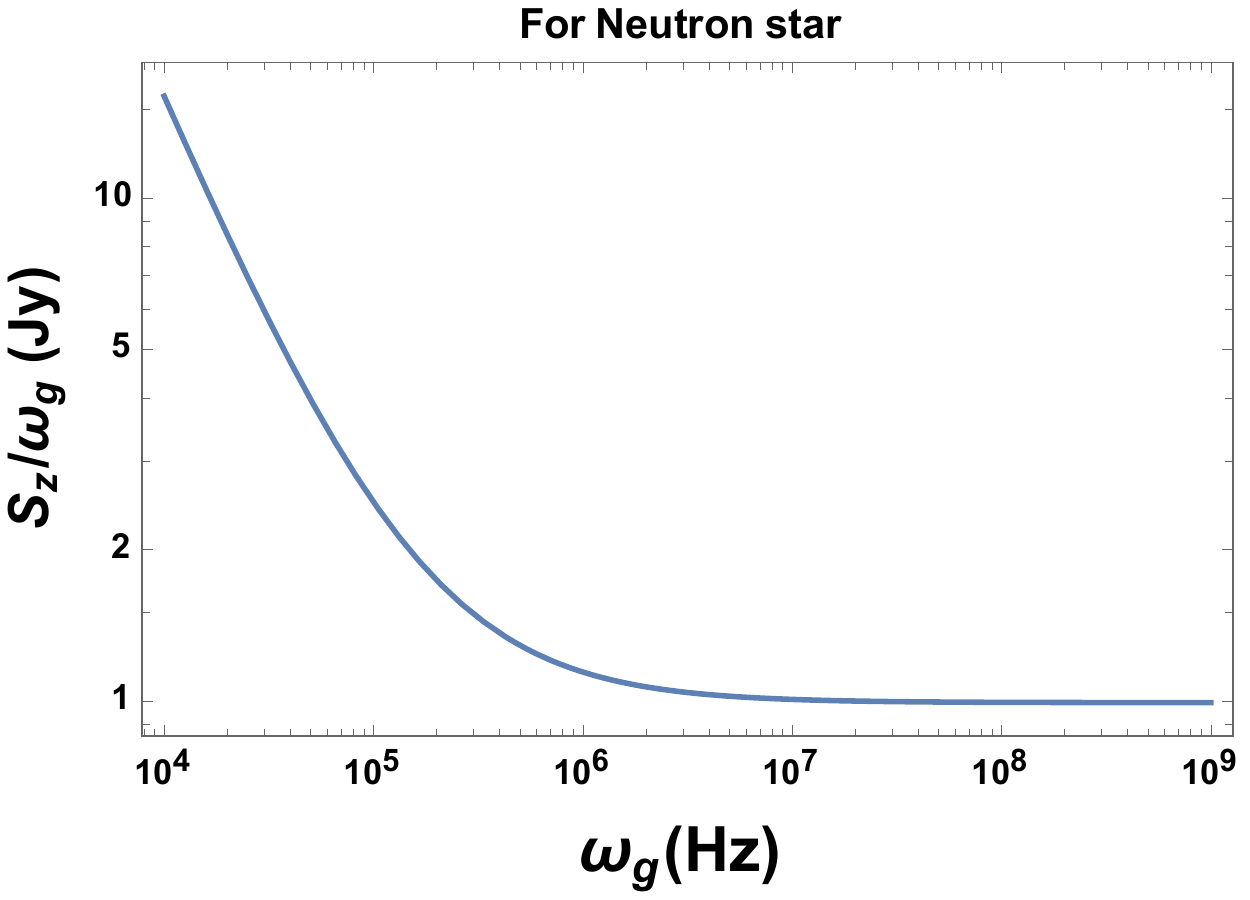} %
\caption{Log-Log plot of $S_z/\omega_g$ versus $\omega_g$ for typical magnetar (top plot) and typical NS (bottom plot). For magnetar, we have set $B^{(0)}_y = 10^{15}  {\rm G}, R_{\rm{LC}} = 10^9~ \, {\rm cm},  \omega_B = 1  \rm{Hz}$. For NS/milli-second pulsar, we have set $B^{(0)}_y = 10^{10} \, {\rm G}  , R_{\rm{LC}} = 10^7 {\rm cm} , \omega_B = 1 \, \rm{kHz}$. For both plots we have set  $A_{+} = 1.4\times 10^{-23}$ corresponding to a typical GW source Aggarwal et al. (2021).}
\label{fig:PoyntVec-Jy}
\end{figure}

Further, to compare the generated EM energy density in the entire magnetosphere with the observations, we compute the flux of the induced EM waves (\ref{eq:E_x-Final}, \ref{eq:B_y-Final}) by calculating the Poynting vector~\cite{1979-Rybicki.Lightman-Book}. 
Rewriting Eq.~(\ref{eq:alpha_tot}) as a quadratic equation in $\xi R_{\rm LC}/c$ provides the functional dependence in-terms of $\alpha_{\rm tot}$. This leads to:
{\small
\begin{align}\label{eq:PoyntingVec-in-alpha}
    S_z \simeq \frac{A_+^2 |B_y^{(0)}|^2 c }{256 \pi} 
    \left[ \sqrt{\frac{24  c^2  \omega_g^2 \alpha_{\rm tot}}{ \pi G | B^{(0)}_y |^2 } - 15} - \frac{12 c^2 \omega_g  \omega_B \alpha_{\rm tot}}{\pi G | B^{(0)}_y |^2 } - 1\right] .
\end{align}
}
Note that $\alpha_{\rm{tot}}$ in RHS of the above expression is a function of $\omega_g$ and the parameters of magnetar/NS. See appendix~\ref{appsec:poyntingVec} for details.
Thus, the Poynting vector of the induced EM waves in the vicinity of magnetar/NS can be obtained by substituting the 
the parameters of magnetar/NS, $\omega_g$ and $A_+$. 
To compare with the peak flux of FRBs, \ref{fig:PoyntVec-Jy} contains the plot of the Poynting vector per unit frequency $(S_z/\omega_g)$ as a function of $\omega_g$ for magnetar and NS. 

The last column in Table \eqref{table1} contains the spectral flux density (Poynting vector per unit frequency) for generic parameter ranges of $R_{\rm LC}$ and $B_{y}^{(0)}$. Given the parameters listed in the table \eqref{table1}, our model predicts a range of spectral flux density that can be as small as $0.1~{\rm Jy}$ (milli-second Pulsar) or  {$10^{-24} {\rm erg ~ s^{-1} cm^{-2} Hz^{-1} }$} and can be as large as $10^{11}~{\rm Jy}$ (Magnetar) or  {$10^{-12} {\rm erg ~ s^{-1} cm^{-2} Hz^{-1} }$}. [ {1 Jy in CGS units is $1 \rm{Jy} = 10^{-23} erg ~ s^{-1} cm^{-2} Hz^{-1}$}.].  We are now in a position to compare the results of the model with radio observations in MHz-GHz frequency. 

\section{A plausible explanation for FRBs}

As mentioned earlier, more than 600 (non-repeating) FRBs are reported in various catalogues~\cite{Petroff:2016tcr,2019-Platts.etal-PhyRept, Pastor-Marazuela:2020tii,2021-Rafiei.etal-APJ}
$99\%$ of these FRBs were found to have the following three characteristic features: peak flux ($S_{\nu}$) varies in the range $0.1~{\rm Jy} < S_{\nu} < 700~{\rm Jy}$, the pulse width is less than one second and coherent radiation~\cite{2021-Rafiei.etal-APJ,Petroff:2016tcr}. In Ref.~\cite{2016-Petroff.etal-PASA}, the authors classified the quantities associated with FRBs into two types --- observed and derived. Peak flux and width are observed quantities, while Luminosity and Luminosity distance are derived quantities. The uncertainty in the Dispersion Measure versus redshift relation can have a cascading effect on the derived quantities of the FRBs, leading to uncertainty in the origin of these events~\cite{2019-Pol.etal-arXiv}. Hence, to reduce the systematic bias, we focus on observed Peak flux and estimate the same using the Poynting vector of the emitted EM waves ($S_{z}/\omega_{g}$). 

%
\begin{table}
\caption{The table contains numerical values of the total conversion factor ($\alpha_{\rm tot}$), energy density of EM waves ($\rho_{\rm EM}$), and spectral flux density (Poynting vector per unit frequency). The first two rows are for a typical Magnetar and the last three rows are for a typical NS.  We have set $A_{+} = 10^{-23}$ corresponding to a typical GW source Aggarwal et al. (2021). Note that if we use $\xi \equiv \delta B_y/B_y^{(0)} \sim 0.1$, $\alpha_{\rm tot}, \rho_{\rm EM}$, and the value of Poynting vector will increase by two-orders of magnitude.}
\label{table1}
\begin{tabular}{|c|c|c|c|c|c|}
\hline
$R_{\rm LC }$   &  $B_y^{(0)}$    &   $\omega_g $    &  $\alpha_{\rm tot}$  & $\rho_{\rm EM} \,  $ & $\frac{S_z}{\omega_g}$ \\ 
(cm)  &  (Gauss)  &  (MHz) &  &  ($ {\rm{erg \, cm^{-3}}}$)  &  (Jy)  \\
\hline
   $10^9$ &  $10^{15}$  & $1$  & $1.74 
   \times 10^{-5}$ & $ {9.3 \times 10^{-13}} $ & $9.95\times 10^{11}$  \\ 
 \hline
   $10^9$  &  $10^{12}$  & $500$  &  $ 1.72 \times 10^{-11}$  & $ {2.3 \times 10^{-13}  }$ & $9.94\times 10^{5}$  \\ 
 \hline
     $10^8$  &  $10^{11}$  & $1400$  &  $ 1.72 \times 10^{-15}$  & $  { 1.8 \times 10^{-16}  }$ & $ 961.57 $  \\ 
 \hline
   $10^7$  &  $10^{10}$  & $1400$  &  $1.72 \times 10^{-19}$  & $  { 1.8 \times 10^{-20} }$ & $0.99$  \\ 
 \hline
   $10^8$  &  $10^{9}$  & $1400$  &  $1.72 \times 10^{-19}$ &  $  { 1.8 \times 10^{-20} }$ & $0.09$  \\ 
 \hline
\end{tabular}
\end{table}

As mentioned earlier, $R_{\rm LC}$ of a 
typical NS is 
$\sim 10^{7}-10^{9}~\rm{cm}$ and it takes less than one second for the GWs to pass through the entire magnetosphere. This directly implies that the induced EM waves due to GZ-effect will appear as a burst lasting for less than one second. 
Thus, our model provides a natural explanation for FRBs lasting less than a second. 
Further, we see from the last column of Table \eqref{table1} that our model predicts the burst of EM wave with the flux $< 1000~\rm{Jy}$. Thus, our model naturally explains the observed peak flux and pulse width of $99\%$ of the reported FRBs.

Having estimated the peak flux and pulse width for our model, we can now calculate other quantities associated with FRBs. In particular, we calculate two such quantities --- Fluence~(${\cal F})$~\cite{2015-Keane.Petroff-MNRAS,2018-Macquart.Ekers-MNRAS} and Isotropic Equivalent Luminosity (IEL)~\cite{2019-Lu.Kumar-MNRAS}. The Fluence is defined as the product of the burst width $\Delta t = (2R_{LC})/c$ and the peak flux ($S_z/\omega_g$)~\cite{2015-Keane.Petroff-MNRAS,2018-Macquart.Ekers-MNRAS}:
\[
{\cal F} = (S_z/\omega_g) \times \Delta t .
\]
Using $S_z$ from the table \ref{table1}, we see that our model predicts the Fluence in the range $1 < {\cal F} ({\rm Jy~ms}) < 10^7$.
We now rewrite the energy density $(\rho_{\rm EM})$ of emitted EM waves in terms of the isotropic equivalent luminosity ($L_{\rm IEL}$). Assuming that the progenitor is at a distance $d$ from the observer, $L_{\rm IEL}$ is given by~\cite{2019-Lu.Kumar-MNRAS}: 
\begin{align}\label{eq:luminosity}
    \rho_{\rm EM} = \frac{L_{\rm IEL}}{4\pi d^2 c}.
\end{align}
The above expression assumes that $d \gg$ size of the progenitor is justified for the reported FRBs. Due to the lack of confirmed detection in other wavelengths, the distance of the FRBs is not well constrained~\cite{2022-Zhang-arXiv}. The dispersion measure of the FRB gives one indirect estimation \cite{2013-Thornton.etal-Science,2016-Ravi.etal-Science}. It is expected that the distance of the FRBs is in the range $[10~{\rm kpc} - 1~{\rm Gpc}]$.  

From the FRB catalogs~\cite{Petroff:2016tcr,2019-Platts.etal-PhyRept, Pastor-Marazuela:2020tii,2021-Rafiei.etal-APJ}, we see that FRBs have a peak flux of around $100 ~{\rm Jy}$. For our model, this corresponds to the energy density of the emitted EM waves to   {be $\sim 9.2 \times 10^{-18}~{\rm erg \, cm}^{-3}$}. Taking the above distance range ($[10~{\rm kpc} - 1~{\rm Gpc}]$), the above energy density translates to $L_{\rm IEL}$ in the range 
$[3.1 \times 10^{39} - 10^{49}]~{\rm erg/s}$. Thus, our model can explain the inferred luminosity of FRBs~\cite{2017-Law.etal-ApJ}.

Given this, we can now identify the possible progenitors of FRBs. To identify this, we consider two FRBs --- FRB120127~\cite{FRB120127} and FRB011025~\cite{FRB011025}. These two sources represent a typical FRB in the catalogs. The observations of these two FRBs at $1.5 ~\rm{and} ~1.3~\rm{GHz}$, show a typical peak flux of $0.62^{+0.35}_{-0.10}~\rm {Jy}$ and $0.54^{+0.11}_{-0.07}~\rm {Jy}$, respectively.
From the last row of the Table \eqref{table1}, we see that our model predicts the progenitor should be a millisecond pulsar with an effective magnetic field strength of $10^{10}~{\rm G}$ and $R_{\rm{LC}} \sim 10^{7}~\rm {cm}$.  {Fluence for these two FRBs is around $0.66~\rm{Jy \, ms}$ and isotropic equivalent luminosity $(L_{\rm IEL})$ at a distance ${d = 10~{\rm Mpc}}$ is $6\times 10^{42} \rm{erg \, s^{-1}}$}.

While many FRBs are considered to be extragalactic, recently, there have been confirmed galactic FRBs~\cite{Bochenek2020Natur,2020Natur.587...54C}, \cite{2019-Lu.Kumar-MNRAS}.
Specifically, FRB 200428 is confirmed to be a galactic FRB~\cite{Bochenek2020Natur}. Recent observations indicate that the magnetar SGR 1935+2154 residing in the Milky Way is associated with the FRB 200428 with a fluence of $>1.5 \times 10^6~{\rm Jy ~ms}$ in the $1.28$–$1.4$ GHz band detected by the STARE2 radio array~\cite{Bochenek2020Natur,2020Natur.587...54C}. It is reported that this magnetar has a surface dipole magnetic field of $\sim 2.2 \times 10^{14}$ G, which can be deduced from the period slow-down rate of $3.24~{\rm s}$. 
Given these observational quantities, we can confirm whether our model can explain the origin of this galactic FRB. To do this, we substitute the constant magnetic field approximation $B_y^{(0)} \sim 2.2 \times 10^{14}~{\rm G}$ with $R_{\rm LC} \sim 10^9~{\rm cm}$ and $A_+ \sim 10^{-26}$ in Eq.\eqref{eq:PoyntingVec-in-alpha}, our model predicts peak flux $\sim 4.8 \times 10^4~{\rm Jy}$ and Fluence to be $3.2 \times 10^6 {\rm Jy~ms}$. Thus our model predicts isotropic equivalent luminosity at $~10 \ {\rm kpc}$ to be $L_{\rm IEL} \sim 2.8\times 10^{43} \rm{erg \, s^{-1}}$. From these, it appears that our model confirms that the FRB 200428 is a galactic burst and originated from the GZ mechanism. Consequently, our model has the potential to explain these observations and play a crucial role in any future FRB and progenitor association (if any).


We can do a similar analysis for 
all the FRBs in the catalog with a pulse width of less than one second~\cite{2021-Rafiei.etal-APJ,Petroff:2016tcr}. Our model predicts that the progenitor can be a NS with an effective magnetic field strength in the range $10^{9} - 10^{11}~{\rm G}$ and rotation frequency $1 < \omega_B < 1000$.  {Our model can provide an explanation for the observed peak flux for a class of \emph{non-repeating} FRBs and predicts FRBs with extremely low and high peak flux. Note that our model is not sensitive to the galactic environment. Such future detections will further strengthen the model.} 

 {As mentioned earlier, the model uses the GZ mechanism to explain the origin of FRBs. GZ mechanism involves the transfer of energy from the incoming GWs to emitted EM radiation in the presence of the background magnetic field. Although our model falls in the first category where FRBs are generated from the magnetar/NS, electromagnetic radiation is generated when GWs pass through the magnetosphere. In other words, the background magnetic field acts as a catalyst in this mechanism.} 
 {We can understand the energy budget of this process by looking at it in two different ways: 
\begin{enumerate}
\item We evaluate the ratio of the energy density of the outgoing EM waves ($\rho_{\rm EM}$) with the energy density of the background magnetic field ($\rho_{B}$). For a magnetic field strength of $10^{10} {\rm G}$, we get 
\[
\frac{\rho_{\rm EM}}{\rho_B} \sim 2.2 \times 10^{-39} .
\]
\item Second, the ratio of the incoming GWs to the emitted EM waves. Rewriting the energy density of the incoming  GWs in terms of the Luminosity and energy density of the emitted EM waves in terms of the Luminosity, we have:
\begin{equation}
L_{\rm EM} \sim \alpha_{\rm tot} L_{\rm GW} \left(\frac{R_{LC}}{D}\right)^2,
\end{equation}
where $L_{\rm GW}$ is given by~\cite{Book-Schutz,2009-Sathyaprakash.Schutz-LivRevRel,2017-LIGOScientific.AnnalPhy}: 
\[ L_{\rm GW} = L_{0} \left( \frac{\omega_g D h}{c} \right)^2 \]
where $L_0 \sim 10^{59}~{\rm erg/s}$ is referred to as fundamental luminosity~\cite{2009-Sathyaprakash.Schutz-LivRevRel}, $\omega_g$ is the frequency of the GWs, and $h$ is the GW amplitude. The detected BH mergers in the LIGO-VIRGO band are of the order $\sim 10^{56}$ erg/s~\cite{2016-LIGOScientific-PRL}. Therefore it is certainly possible to have such large luminosity GW sources. In fact, for highly-relativistic systems,  $L_{\rm GW} \sim L_0$\cite{2009-Sathyaprakash.Schutz-LivRevRel}. \\
From the above two expressions and taking the values we have used in the earlier section, we see that our model predicts 
$L_{\rm EM} \propto 10^{36}~{\rm erg/s}$. However, if we use $\xi \equiv \delta B_y/B_y^{(0)} \sim 0.1$, our model predicts $\alpha_{\rm tot} \sim 10^{-3}$ and can lead to larger value of $L_{\rm EM}$. \\
Note that $L_{\rm EM}$ derived is theoretical Luminosity based on the Luminosity of the incoming GW and is not equivalent to $L_{\rm IEL}$ defined in Eq.~\ref{eq:luminosity}. From the above discussions, it s clear that our mechanism is energetically expensive and may not be a frequent phenomenon.
\end{enumerate}
}

\section{Discussions}

Our mechanism requires that the emitted GWs pass through the NS, resulting in \emph{non-repeating} FRBs along the line of sight of the observer. In other words, we have assumed that the EM emission due to the GZ-mechanism in the pulsar magnetosphere is directional dependent. The intervening medium does not impact the radiation from this mechanism. More importantly, the mechanism requires that the emitted GWs pass through the NS, resulting in FRBs along the line of sight of the observer. The probability of this event is a product of the probability that the GW passes through the NS and the probability that the emitted EM is along the line of sight of the observer. Hence, the probability of seeing such an event is not very high. More importantly, any existing NS/Magnetar in any galaxy can produce FRB. Interestingly, recently GZ effect was employed by  \cite{Kalita:2022uyu} to compare the expected output of the GZ signal with two known FRBs and their potential detection in GW detectors. They demonstrated that any continuous GW signal detected by the suggested GW detectors from the FRB location would instantly suggest that the merger-like theories are unable to account for all FRBs, hence offering strong evidence in favor of the GZ hypothesis.


Further, it is estimated that around $10^8 - 10^9$ NSs are present in the Milky Way galaxy, roughly $1\%$ of the total number of stars in the galaxy~\cite{2006-Diehl.etal-Nature,2010-Sartore.etal-AandA}. Also, it is estimated that the magnetar formation rate is approximately $1 - 10$ percent of all pulsars~\cite{2015-Gullon.etal-MNRAS,2019-Beniamini.etal-MNRAS}.   
One of the basic assumption of our model is that, given GW signal in MHz-GHz frequency, all NSs can act as source of FRBs at all times. This assumption translates to the fact that the maximum FRB events per day can be $10^8 - 10^9$.  However, the observed FRB rate is $10^{3}$ for the entire sky per day. This can be attributed to the fact that the probability of this event is a product of the probability that the GW passes through the NS times the probability that the emitted EM is along the line of sight of the observer.
%
Consequently, our model predicts the observed FRB event rate and the coherent nature of FRBs~\cite{2013-Katz-PRD}. Since, GWs can be generated up to 14~GHz~\cite{2022-Aggarwal.etal-PRL,2020-Ito.etal-EPJC}, our model naturally does not have high-energy counterpart. 

As mentioned in the Introduction, the co-planar property of the EM wave emitted due to the GZ mechanism is due to the coherent nature of the incoming GWs~\cite{2007-Hendry.Woan-AG}. 
Due to the co-planar property, it can maintain the flux~\cite{2022-Lieu.etal-CQG}. Hence, the three key features of non-repeating FRBs are naturally explained in our model. Extending the calculation for repeating will be discussed in future work.

A variety of processes generate GWs in a broad range of frequencies~\cite{1974-Hawking.Carr-MNRAS,2009PhRvL.103k1303A,2015-Kuroda.etal-IJMPD,2019-Ejlli.etal-EPJC,2020-Aggarwal.etal-arXiv,2021-Pustovoit.etal-JOP}. However, it is possible to detect these waves only in a limited range. The proposed model provides an indirect mechanism to detect GWs at high frequencies. Interestingly, our model also provides a way to test modified theories of gravity. In this work, we have focused on GWs in general relativity. Certain modified theories like Chern-Simons gravity lead to birefringence~\cite{2006-Alexander.Peskin.Jabbari-PRL} which can explain the polarized nature of FRBs. However, exact rotating solutions in these theories are unknown and require sophisticated numerics. Thus, the high-frequency GWs will provide a unique view of the Universe --- it's birth and evolution.

\section*{Acknowledgments}
We thank the referee for the valuable comments that significantly improved the manuscript. The authors thank Sushan Konar and Surjeet Rajendran for discussions. The authors thank Koustav Chandra, S. M. Chandran, Archana Pai, and T. R. Seshadri for their comments on the earlier draft. AK thanks L. F. Wei for clarifications of reference. AK is supported by the MHRD fellowship at IIT Bombay. This work is supported by ISRO Respond grant.

\section*{Data Availability}
We have not utilized any separate dataset for our calculation. The data used for the plots are from the equations given in the paper. 

\bibliography{References}
\bibliographystyle{mnras}

\appendix

\section{Gertsenshtein-Zel$'$dovich effect and induced electromagnetic waves}
\label{appsec:E&B-solution}
%

Consider a background space-time with GWs~\cite{Book-Gravitation_MTW}:
\begin{align}\label{eq:Metric}
g_{\alpha\beta} = \eta_{\alpha\beta} + h_{\alpha\beta} \, , \qquad 
g^{\alpha\beta} = \eta^{\alpha\beta} - h^{\alpha\beta}
\end{align}
where $h_{\alpha\beta} \ll 1$ is the GW fluctuation, and $\eta_{\alpha\beta} = \rm{diag}(1,-1,-1,-1)$ is Minkowski space-time. The background space-time is generally curved; however, the results derived for Minkowski space-time carry through for the conversion factor computations. For the generic curved background, the Riemann corrections contribution is tiny~\cite{Book-Gravitation_MTW}. Hence, we only report the results for the Minkowski space-time. 

For the GWs propagating in the $z-$axis, we have $h_{xx} = - h_{yy} = h_{+}$, $h_{xy} = h_{\times}$. 
%
%
Eq. \eqref{eq:h-Expression} corresponds to a monochromatic circularly polarized GW~\cite{Book-Gravitation_MTW,2018-Zheng.Wei.Li-PRD}. As mentioned earlier, the presence of magnetic field transverse to the direction of propagation of GWs acts as a catalyst for the conversion process \cite{1974-Zeldovich-SJETP}. As discussed, we consider the total magnetic field in the magnetosphere to be
\begin{align}\label{eq:bg-magnetic_field}
\textbf{B}(t) = \left( 0, B^{(0)}_y + \delta B_y \sin (\omega_B t), 0  \right)
\end{align}
where $B^{(0)}_y$ is static magnetic field and $ B_y \sin (\omega_B t)$ is the alternating (time-varying) magnetic field with frequency $\omega_B $ and $| \delta B_y| < |B^{(0)}_y| $.
Also, we assume that the amplitude of the alternating magnetic field is two orders lower than the static magnetic field, i.e., 
$| {B_y}/{B^{(0)}_y}| \approx 10^{-2}$, so that time-varying magnetic field has significant effects on the conversion~(\cite{2012-Pons.etal-AandA,2019-Pons.Vigan-arXiv}). The induced electric field due to the time-varying magnetic field is~[\cite{Book-Jackson-Classical_Electrodynamics}]
\begin{align}\label{eq:InducedElectricField}
\textbf{E}(z,t) = \left( - \frac{z\, \delta \omega_B B_y}{c} \cos(\omega_B t)  , 0 , 0 \right) \, .
\end{align}
%
%
In the presence of GWs, the EM field tensor is:
\begin{align}\label{eq:EMField-tensor-dd_matrix}
F_{\alpha\beta} = F^{(0)}_{\alpha\beta} + F^{(1)}_{\alpha\beta} 
= \begin{pmatrix}
0 & \mathcal{E}_x & \tilde{E}_y & \tilde{E}_z \\
-\mathcal{E}_x & 0 & -\tilde{B}_z & \mathcal{B}_y \\
-\tilde{E}_y & \tilde{B}_z & 0 & -\tilde{B}_x\\
-\tilde{E}_z & - \mathcal{B}_y & \tilde{B}_x & 0  
\end{pmatrix}  ,
\end{align}
where $F^{(1)}_{\alpha\beta}$ is the field tensor induced due to the GWs that needs to be determined. Similarly, the induced electric 
$[\tilde{ \textbf{E} } = \left(  \tilde{E}_x, \tilde{E}_y, \tilde{E}_z \right)]$ 
and magnetic field vectors  $[\tilde{\textbf{B} }= \left( \tilde{B}_x, \tilde{B}_y, \tilde{B}_z \right)$] due to GWs are to be determined.
Note that $\mathcal{B}_y = B^{(0)}_y + \delta B_y \sin (\omega_B t) + \tilde{B}_y $ and $ \mathcal{E}_x = - (z \, \delta B_y \omega_B/{c}) \cos(\omega_B t) + \tilde{E}_x$. 
The covariant Maxwell's equations (in the source-free region) are:
%
\begin{align}
\label{eq:MaxwellEq12}
\partial_{\mu} \left( \sqrt{-g} F^{\mu\nu} \right) = 0;~~
 \partial_{\mu} \left( \sqrt{-g} \tilde{F}^{\mu\nu} \right) = 0
\end{align}
where $\tilde{F}^{\mu\nu} =  \epsilon^{\mu\nu\alpha\beta} F_{\alpha\beta}/2$ is the dual of EM field tensor, $\epsilon^{\mu\nu\alpha\beta}  = \eta^{\mu\nu\alpha\beta}  /{ \sqrt{-g} }$ and $\eta^{\mu\nu\alpha\beta}$ is defined as $\eta^{0123} = 1 = - \eta_{0123} $ is an antisymmetric tensor.
Substituting Eq.~(\ref{eq:EMField-tensor-dd_matrix}) in Eq.~\eqref{eq:MaxwellEq12}, and treating $F^{(1)}_{\alpha\beta} $ and $h_{\alpha\beta}$ as first-order perturbations, we have
%
\begin{subequations}
\begin{align}
\label{eq:dtEx}
&{} \frac{1}{c} \partial_t \tilde{E}_x - \partial_y \tilde{B}_z + \partial_z \tilde{B}_y 
+ \left( B^{(0)} + \delta B_y \sin(\omega_B t) \right) \, \partial_z h_{+}  
\nonumber\\ 
&{}- \frac{ z \, \delta B_y \omega_B}{c^2} \frac{\partial }{\partial t} \left( h_{+} \, \cos(\omega_B t) \right) = 0 \\
\label{eq:dtBy}
&{}\frac{1}{c} \partial_t \tilde{B}_y - \partial_x \tilde{E}_z + \partial_z \tilde{E}_x  = 0 \, ,
\end{align}
\end{subequations}
where we have expressed in terms of electric and magnetic fields. Since GWs (and EM waves) propagate along the z-direction, we have $\tilde{E}_z = \tilde{B}_z = 0$. Substituting Eq.~(\ref{eq:h-Expression}) in  the above wave equations lead to the following wave equations for $\tilde{E}_x, \tilde{B}_y$:
\begin{subequations}
\begin{align}\label{eq:WaveEquation-Ex-FF}
\frac{1}{c^2} \frac{\partial^2  \tilde{E}_x }{\partial t^2} - \partial_z^2 \tilde{E}_x  
&= - f_{E}(z^{\prime},t^{\prime}) \\
\label{eq:WaveEquation-By-FF}
\frac{1}{c^2} \frac{\partial^2  \tilde{B}_y }{\partial t^2} - \partial_z^2 \tilde{B}_y 
&= - f_{B}(z^{\prime},t^{\prime})
\end{align}
\end{subequations}
where $f_{E/B}(z^{\prime},t^{\prime})$ are the forcing functions and are given by: 
\begin{subequations}
\begin{align}\label{eq:ForcingFunction-E}
&f_{E}(z^{\prime},t^{\prime}) =  
\frac{ i A_{+} \delta B_y k_g }{ 2 c}  
\left[ \omega_{+}  e^{i ( k_g z^{\prime} - \omega_{+} t^{\prime}) } - \omega_{-} 
e^{i  ( k_g z^{\prime} - \omega_{-} t^{\prime})} \right]  
\nonumber\\
& + \frac{ z^{\prime} A_{+} \delta B_y \omega_B }{ 2 c^3}  \left[\omega_{+}^2 e^{i ( k_g z^{\prime} - \omega_{+} t^{\prime} ) } + \omega_{-}^2 e^{i ( k_g z^{\prime} - \omega_{-} t^{\prime}) }  \right]  \nonumber \\
& + \frac{A_{+} B^{(0)} k_g \omega_g  }{c}  
e^{i ( k_g z^{\prime} - \omega_g t^{\prime}) } \\
\label{eq:ForcingFunction-B}
& f_{B}(z^{\prime},t^{\prime}) = 
\frac{ i A_{+} \delta B_y k_g^2  }{ 2 } \left[ e^{i \left( k_g z^{\prime} - \omega_{+} t^{\prime} \right) }  - e^{i \left( k_g z^{\prime} - \omega_{-} t^{\prime} \right) } 
\right]
\nonumber\\
& + \frac{ i A_{+} \delta B_y \omega_B   }{ 2 c^2}    \left[ \omega_{+}  e^{i \left( k_g z^{\prime} - \omega_{+} t^{\prime} \right) }  -  \omega_{-}  e^{i \left( k_g z^{\prime} - \omega_{-} t^{\prime} \right) } 
\right]
\nonumber\\
& -  \frac{ z^{\prime} A_{+} \delta B_y \omega_B k_g   }{ 2 c^2}  \left[ \omega_{+}  e^{i( k_g z^{\prime} - \omega_{+} t^{\prime}) }   - \omega_{-}  e^{i( k_g z^{\prime} - \omega_{-} t^{\prime}) }  \right] \nonumber \\
& + A_{+} B^{(0)} k_g^2   e^{i \left( k_g z^{\prime} - \omega_g t^{\prime} \right) } 
\end{align}
\end{subequations}
The solutions to the wave equations (\ref{eq:WaveEquation-Ex-FF}) and (\ref{eq:WaveEquation-By-FF}) are given by: 
\begin{align}\label{eq:ResponseToForcingFunction}
F_{E/B}(z,t) = \int \int dz^{\prime} dt^{\prime}  G( t,t^{\prime} ; z,z^{\prime} )  f_{E/B}( t^{\prime},z^{\prime} ) 
\end{align}
where $F_{E/B}(z,t)$ denotes the corresponding solution of the forcing function $f_{E/B}(z^{\prime},t^{\prime})$ and $G( t,t^{\prime} ; z,z^{\prime})$ is the retarded Green's function corresponding to the wave equation and is given by:
\begin{align}\label{eq:GreenF}
G( t,t^{\prime} ; z,z^{\prime} ) = \frac{c }{2} \Theta \left( \,\, c(t-t^{\prime}) - |z-z^{\prime}| \,\, \right) 
\end{align}
where $\Theta-$function is non-zero only for $t \geq  t^{\prime} + \frac{|z-z^{\prime}|}{c}$. Substituting the forcing functions for the electric (\ref{eq:ForcingFunction-E}) and the magnetic (\ref{eq:ForcingFunction-B}) fields, in the integral equation (\ref{eq:ResponseToForcingFunction}), leads to:
\begin{align}\label{eq:E_x-Expression}
\tilde{E}_x &\simeq  -\frac{B^{(0)}_y A_{+} }{4} \,\, e^{i \left(k_g z - \omega_g t\right) }  
\\
& +  \frac{ \delta B_y A_{+}  \omega_B t }{2} \,\, e^{i \left(k_g z - \omega_g t\right) } 
+  \frac{ \delta B_y A_{+} }{4 i } \, \left( \frac{\omega_B}{\omega_g} \right) \,\, e^{i \left(k_g z - \omega_g t\right) }  
\nonumber \\
\label{eq:B_y-Expression}
\tilde{B}_y &\simeq  -\frac{B^{(0)}_y A_{+} }{4} \,\, e^{i \left(k_g z - \omega_g t\right) }  \\
& -  \frac{ \delta B_y A_{+}  \omega_g t }{2} \,\, e^{i \left(k_g z - \omega_g t\right) } 
 +  \frac{ \delta B_y A_{+} }{4 i } \, \left( \frac{\omega_B }{\omega_g} \right)^2 \,\, e^{i \left(k_g z - \omega_g t\right) }  
 \,\,  .
 \nonumber 
\end{align}
We want to mention the following points regarding the above expressions: First, in obtaining the above expressions, we have assumed that $ \omega_B \ll {\omega_g}$. This is valid in our case because the conversion from GWs to EM waves occurs in the MHz-GHz frequency range. This approximation leads to $\omega^n_{\pm} \simeq \omega^n_g \left( 1 \pm n \frac{\omega_B}{\omega_g} \right)$ and allows us to approximate $\omega_{\pm} \approx \omega_g$ in the exponentials. Note that we have also ignored the terms with $e^{ i\omega_g t}$ since they will lead to wave propagating along negative z-direction i.e., $e^{i \left(k_g z + \omega_g t\right) }$, and hence are not relevant to our analysis.
Second, the last term in both expressions is tiny and can be neglected leading to Eqs.~(\ref{eq:E_x-Final}, \ref{eq:B_y-Final}). 
 {Lastly, GZ effect is a pure gravitational effect (due to incoming GWs) and the generation of EM waves does not require any source term (plasma or charged particles). Hence, if the incoming GWs are coherent, the emitted EM waves are coherent at resonance.}

\section{Evaluating the radius of the magnetosphere}
\label{appsec:Integration-2}

We show that the assumption that the background magnetic field can be treated as a constant in the entire magnetosphere gives identical results to that of the background field decreasing radially. 

In evaluating $\alpha_{\rm tot}, $ we have assumed that the background magnetic field is a constant until light cylinder radius $R_{\rm LC}$. In this section, we show that this assumption is physically consistent. 

To do that, we consider the magnetic field of the pulsar magnetosphere in vacuum to be dipolar. 
We evaluate the average dipolar magnetic field in the magnetosphere. In spherical polar coordinates, the dipolar magnetic field is given by~\cite{2016-Cerutti.Beloborodov-SpaceSciRev,2016-Petri-JPlasmaPhy}:
%
\begin{align}\label{eq:B-stellar_equator}
\left(  B_r , B_{\theta} , B_{\phi}  \right) = B_{*} \, \left( \frac{r_* }{r} \right)^3 \, \left(  2 \cos\theta, \sin\theta , 0  \right)
\end{align}
where $B_*$ is the magnetic field on the NS surface. 

We now compute the average magnetic field on the volume between the radius of the NS ($r_*$) and 
$R_{ \rm{LC}}$:
\begin{equation}\label{eq:Vol_avg_magnetic}
\!\! \bar{ \textbf{B} }(r,\theta,\phi) = 
\frac{1 }{V}
\int_{r_{*} }^{ R_{ \rm{LC} } }  \!\!\!\!\!\! dr  \int_0^{\pi}  \!\!\!\! d\theta \int_0^{2 \pi } \!\!\!\! d\phi \, r^2 \sin\theta \, \textbf{B} (r, \theta, \phi) 
\end{equation}
where $V$ is the volume enclosed by both the surfaces:
\begin{align}\label{Volume}
V = \frac{4 \pi}{3} \left(   R_{ \rm{LC} }^3  - r_{*}^3  \right).
\end{align}
From Eqs.~(\ref{eq:Vol_avg_magnetic}, \ref{Volume}), we define the integral $\mathcal{I}$ as: 
{\small
\begin{align}\label{eq:Integration-I-Def}
\!\!\!\! \mathcal{I} \equiv V \, \bar{ \textbf{B} }(r,\theta,\phi)  =  
\int_{r_{*} }^{ R_{ \rm{LC} } } \!\!\!\! dr \int_0^{\pi} \!\!\!\! 
d\theta \int_0^{2 \pi } \!\!\!\! d\phi \, r^2 \sin\theta 
\textbf{B} (r, \theta, \phi) .
\end{align}
}
Note that the volume average of the radial component of the magnetic field vanishes.
The non-zero contribution comes from the angular variation of the magnetic field, i.e.,
\begin{align}\label{eq:Vol_avg_magnetic-theta}
\bar{\textbf{B}}_{\theta}  &= \frac{3 \pi B_* r_{*}^3 }{4 \left(  R_{ \rm{LC} }^3 - r_{*}^3  \right) } \ln \left| \frac{ R_{ \rm{LC} } }{ r_* }  \right| .
\end{align}
From Eqs.~(\ref{eq:Integration-I-Def}, \ref{eq:Vol_avg_magnetic-theta}), we obtain
\begin{align}\label{eq:I-eval-1}
\mathcal{I} &= 
 \pi^2 B_* r_{*}^3  \ln \left| \frac{ R_{ \rm{LC} } }{ r_* }  \right|
\end{align}
We now compare this with the assumption that the 
magnetic field is approximately constant until $R_{\rm LC}$. To do this, we substitute the 
average magnetic field obtained in Eq.~(\ref{eq:Vol_avg_magnetic-theta}) inside the integral in  Eq.(\ref{eq:Integration-I-Def}). 
Doing the radial integral between $r_*$ to  $\mathcal{R}$ leads to:
\begin{align}\label{eq:I-eval-2}
\mathcal{I} &=  
\pi^2 B_* r_{*}^3 \frac{ ( \mathcal{R}^3 - r_*^3 ) }{ \left(  R_{ \rm{LC} }^3 - r_{*}^3  \right) }  \,   \ln \left| \frac{ R_{ \rm{LC} } }{ r_* }  \right|.
\end{align}

Setting $\mathcal{R} = R_{ \rm{LC} } $ in the above expression, we see that the two expressions (\ref{eq:I-eval-1}, \ref{eq:I-eval-2}) are approximately the same. Thus, even if there is a radial and angular variation in the magnetic field in the magnetosphere, we can approximate the background field to be approximately constant in the region $ r_* \leq r \leq R_{ \rm{LC} }$. Thus, our expression for the total conversion factor mimics the realistic NS regions. 

\section{Conversion factor from entire magnetosphere}
\label{appsec:Integration}
In this appendix, we estimate the total conversion factor from the entire magnetosphere of the compact object. 
The energy density carried by these induced EM waves is~\cite{Book-Jackson-Classical_Electrodynamics}
\begin{align}\label{eq:rho_EM}
\rho_{\rm{EM} }  
\simeq \frac{ |A_{+} |^2 |B^{(0)}_y|^2 }{64 \pi}  \left[ 1   + 2 \, \xi^2 \omega_g^2 \, t^2 + 2 \, \xi \omega_g \, t  \,\,  \right] \, .
\end{align}
where $\xi = \delta B_y/B_y^{(0)}$. As mentioned earlier, the conversion is maximum when EM waves are approximately the same as the incoming waves. 
Having obtained the energy carried by the induced EM waves, we need to obtain what fraction of wave energy is converted to EM waves~\cite{1974-Zeldovich-SJETP}? To do this, we calculate the energy density carried by the GWs~\cite{Book-Gravitation_MTW}, i. e.:
\begin{align}\label{eq:rho_GW}
\rho_{\rm{GW} } = \frac{c^2 \omega_g^2}{32 \pi G } \left(  |A_{+}|^2 + |A_{\times}|^2 \right) = \frac{c^2 \omega_g^2}{16 \pi G } \,\,  |A_{+}|^2 \, ,
\end{align}
wherein second equality, we have used the fact that both the modes of GWs are generated with an equal amount of energy, i.e., $|A_{+}| = |A_{\times}| $ also referred to as isospectrality relation~\cite{Chandrasekhar_BlackHoles-Book}. From Eqs.~(\ref{eq:rho_EM}, \ref{eq:rho_GW}), we obtain Eq.~\eqref{eq:alpha-at-point}. 

To obtain the conversion factor in the entire magnetosphere, we define the following dimensionless parameter:
\begin{align}\label{eq:define-X}
X = \frac{r}{R_{ \rm{LC} }}
\end{align}
where $r$ is the distance from the surface of neutron star to a point in the magnetosphere, i.e., $r_* \lesssim \,  r  \lesssim  \, R_{\rm{LC}}$, and $r_* (= 10^6 \rm{cm})$ is the radius of the Neutron star (NS). In the case of Magnetar $R_{\rm LC} = 10^9~{\rm cm}$ and for NS $R_{\rm LC} = 10^7~{\rm cm}$, hence, $X_* = \frac{r_*}{R_{\rm{LC}} } < 1$. In other words, the range of $X$ is $ 0.001 \lesssim \,   X   \lesssim \,  1 $. 

Since we are interested in the EM waves reaching the observer, we are interested in evaluating the conversion factor along the observer's line of sight. Thus, we have 
\begin{align}\label{eq:z=XR}
z = r = XR_{\rm{LC}}.
\end{align}
where $\theta = 0$ corresponds to the direction along the line-of-sight of the observer.  Thus, the total conversion factor is given by the integral
\begin{align}\label{eq:alpha_tot-def}
\alpha_{\rm{tot} } = \Omega \int_{X_*}^{1} \alpha \,\, dX  ,
\end{align}
where $\Omega$ is the total solid angle which is an overall constant factor because Eq.(\ref{eq:alpha_tot-def}) is independent of the angular coordinates.

\section{Poynting vector}
\label{appsec:poyntingVec}

In astrophysical observations of compact objects, a quantity of interest is the energy flux density which is the Poynting vector. This section evaluates the Poynting vector for magnetar and neutron star in SI unit and Jansky Hz, a widely used unit for spectral flux density in radio observations.

The Poynting vector of the induced electric field (\ref{eq:E_x-Final}) and induced magnetic field (\ref{eq:B_y-Final}) is~\cite{Book-Jackson-Classical_Electrodynamics}:

\begin{align}\label{eq:poyntingVector}
    S_z &= \frac{c}{8 \pi} \tilde{E}_x \times \tilde{B}^*_y \\ 
    & \simeq \frac{A_+^2 \, |B_y^{(0)}|^2  \, c }{128 \pi} \left[ 1 + 2 \omega_g \xi \frac{R_{\rm{LC}}}{c} - 2 \omega_g \omega_B \xi^2 \left( \frac{R_{\rm{LC}}}{c}\right)^2  \right] \nonumber 
\end{align}
where $\tilde{B}^*_y$ is the complex conjugate of the induced magnetic field $\tilde{B}_y$.  As mentioned above, the frequency of the alternating magnetic field is $1 \rm{Hz}$ for the magnetar and $10^3 \, \rm{Hz}$ for the millisecond pulsar. Here again, we have assumed that $\omega_B \ll \omega_g$. Rewriting the above Poynting vector in terms of $\alpha_{\rm{tot} }$, we get Eq.~\eqref{eq:PoyntingVec-in-alpha}. 

\section{Current status of high-frequency GWs}\label{appsec:ECO}
 {
Primordial black-holes, Exotic compact objects and early Universe can generate high-frequency GW (HFGW) in MHz to GHz~\cite{2008-Akutsu.etal-PRL,2008-Nishizawa.etal-PRD,2016-Holometer-PRL,2017-Chou.etal-PRD,2020-Ito.etal-EPJC,2021-Domenech-EPJC,2021-Goryachev.etal-PRL,2022-Aggarwal.etal-PRL,2022-Domcke.etal-PRL}. These can generate GWs up to 14~GHz~\cite{2020-Ito.etal-EPJC}. Over the last decade, many HFGW detectors are proposed, and some of them are operational. For instance, the Japanese 100 MHz detector with a $0.75~{\rm m}$ armlength interferometer has been operational for a decade~\cite{2008-Akutsu.etal-PRL,2008-Nishizawa.etal-PRD}, Holometer detector has put some limit on GWs at MHz~\cite{2016-Holometer-PRL,2017-Chou.etal-PRD}. 
A GHz GW detector is also proposed~\cite{2020-Ito.etal-EPJC}. These detectors are ideally suited for searching for physics beyond the standard model (SM), like primordial black-holes, exotic compact objects and the early Universe. For instance, exotic compact objects with characteristic strain $h$~\cite{2020-Aggarwal.etal-arXiv}:
\[
h \lesssim  10^{-19} C^{5/2} \left( \frac{\rm MHz}{f} \right) \left( \frac{\rm Mpc}{D} \right)
\]
lead to the following coherent GWs with amplitudes: 
 \[
 h_{\rm 1.4 GHz, 10kpc} \lesssim 10^{-21} \, ,
  h_{\rm 1.4 GHz, 1Mpc} \lesssim 10^{-23} \ .
 \]
}
{After 153 days of operation, the Bulk Acoustic GW detector experiment recently reported two MHz events~\cite{2021-Goryachev.etal-PRL}. According to Goryachev et al. ~\cite{2021-Goryachev.etal-PRL}, the data corresponding to two MHz events fits best a single energy depositing event. The authors mention that the strongest observed signal results in a required characteristic strain amplitude of $h_c \approx 2.5 \times 10^{-16}$, corresponding to a PBH merger of $m_{\mathrm{PBH}}<4 \times 10^{-4} M_{\odot}$ (which gives a maximum frequency at inspiral of $5.5 \ \mathrm{MHz}$), at a distance of $D \approx 0.01 \mathrm{pc}$. It is important to note that these are not conclusive enough. However, these detections also imply that these events are not rare.}

 {As mentioned above, many HFGW detectors are proposed, and better estimates will be available as more and more detectors will be operational in the coming decades. This can provide better estimate of these events in the next decade.}

\end{document}